\documentstyle[preprint,prl,aps]{revtex}

\begin{document}
\draft

\title{EXACT SOLUTION OF A ONE-DIMENSIONAL MULTICOMPONENT LATTICE GAS
WITH HYPERBOLIC INTERACTION}

\author{Bill Sutherland and Rudolf A.\ R\"{o}mer\cite{rar}}

\address{Physics Department, University of Utah, Salt Lake City, Utah 84112,
USA}


\author{B. Sriram Shastry}

\address{Physics Department, Indian Institute of Science, Bangalore 560094,
INDIA}

\date{Long Version: April 29, 1994; printed \today}
\maketitle

\begin{abstract}
	We present the exact solution to a one-dimensional multicomponent
quantum lattice model interacting by an exchange operator which falls off as
the inverse-sinh-square of the distance.  This interaction contains a variable
range as a parameter, and can thus interpolate between the known solutions
for the nearest-neighbor chain, and the inverse-square chain.  The energy,
susceptibility, charge stiffness and the dispersion relations for low-lying
excitations are explicitly calculated for the absolute ground state, as a
function
of both the range of the interaction and the number of species of fermions.
\end{abstract}

\pacs{03.65.Db}

\narrowtext

%
%

\section{The Lattice Model from the Exchange Model}

	In a recent Physical Review Letter \cite{1}, to be called S2, Sutherland and
Shastry introduced a one-dimensional, continuum, multi-component
quantum many-body system with a hyperbolic exchange interaction,
governed by the following Hamiltonian:
\begin{equation}
H_{cont}=-{1 \over 2}\sum\limits_j {{{\partial ^2} \over {\partial
x_j^2}}}+\sum\limits_{j>k} {{{\lambda ^2-\lambda P_{jk}} \over {\sinh
^2(x_j-x_k)}}}.
\end{equation}
We have set $\hbar=1$, the mass to be unity, $P_{jk}$ is the exchange operator
which permutes particles $j$ and $k$, and we always have in mind the
thermodynamic limit, when the number of particles $N$ and the box size
$\Lambda$
both become infinite, while the density $N/\Lambda=d$ remains finite.
In S2, scattering in this system was shown to be non-diffractive, and so the
system could be solved exactly (in the thermodynamic limit, with finite size
corrections being exponentially small in $\Lambda$) by the asymptotic Bethe
ansatz
\cite{2}.
The system is characterized by the number of different types b(f) of bosons
(fermions) in the system;  we write this as a $B^bF^f$ system.  Thus $B(F)$
represents a single component system of identical bosons(fermions) interacting
with an interaction strength $\lambda(\lambda \mp 1)$.

This system has a classical limit, obtained by taking
$\lambda\rightarrow\infty$, and in the ground state, the particles
{\em freeze} into a lattice with lattice spacing $\Lambda/N=1/d$,
so that $x_j\rightarrow j/d$.  Expanding the Hamiltonian in $1/\lambda$,
we find $H_{cont}\rightarrow H_{elas} + 2\lambda H_{latt}$, where $H_{elas}$
represents the Hamiltonian for the elastic degrees of freedom --- phonons and
solitons --- identical to the single component $F$ system \cite{3}.
The additional term of order $\lambda$ gives the Hamiltonian for the
compositional degrees of freedom $H_{latt}$,
\begin{equation}
H_{latt}=-{1 \over 2}\sum\limits_{j>k} {{{1+P_{jk}} \over {\sinh ^2[(j-
k)/ d]}}}.
\label{eq-2}
\end{equation}
There is exactly one particle to a site.  Thus, to order $\lambda$, the
elastic and compositional degrees of freedom separate.  Coupling between the
two occurs through an expansion of the interaction about the regular lattice,
and so is of order $\langle\delta x^2 \rangle \sim 1/\lambda$, and down by
another factor of $1/\lambda$.
(A similar procedure is used to derive the Heisenberg antiferromagnet from the
half-filled Hubbard model.)

In this letter, we will study the spectrum of this lattice Hamiltonian.
The Hamiltonian has many realizations --- as does the corresponding
nearest-neighbor model --- such as by spins, a (supersymmetric) $t$-$J$ model,
$SU(N)$ spins, etc. In fact, this system interpolates between the
nearest-neighbor model \cite{4,5} ($d\rightarrow 0$), and the $1/r^2$ lattice
\cite{1,6,7} ($d\rightarrow\infty$) treated in S2.  Our solution
here will rely heavily on Sutherland's treatment \cite{4}
--- to be called S1 --- of the nearest-neighbor system.
Two results proven in S1 can easily be shown for the hyperbolic case:
(i) the ground state of the $B^bF^f$ system is identical to the
ground state of the $BF^f$ system, so for the ground state, we need never
consider more than one type of boson which we usually think of as a vacancy;
(ii)  the Hamiltonian with $P_{jk} \rightarrow - P_{jk}$ is unitarily
equivalent to $H_{latt}$, hence "ferro-" and "antiferro-" cases are both
included in the ground state of $H_{latt}$.
The pure (anti)ferro- case is $B^b$($F^f$).  Partial results concerning the
integrability of this model have previously appeared in the literature
\cite{8}.

%
%

\section{The Solution of the Exchange Model}

Because the system is non-diffractive, the wavefunction must be given
asymptotically as
\begin{equation}
\Psi (x|Q)\sim\sum\limits_P {\Psi (P|Q)}\exp [i\sum\limits_{j=1}^N
{x_jk_{Pj}]},
\end{equation}
where $x_1 < \ldots < x_N$, $P$ is one of the $N!$ permutations of the $N$
asymptotic momenta $k_j$, $Q$ is one of the $N!$ rearrangements of the
particles, and $\Psi(P|Q)$ are $(N!)^2$ amplitudes related by two-body
scattering.  The $k$'s are determined by a set of eigenvalue equations,
which say that the phase shift of a particle going around the ring must be
unity $\mbox{mod}\ 2\pi$:
\begin{equation}
\exp [ik_j\Lambda ]\prod\limits_{m(\ne j)} {{-\exp [-i\theta _0(k_j-
k_m)]}\,}X_{j,j-1}\cdots X_{j,1}X_{j,N}\cdots X_{j,j+1}\,\Psi =\Psi.
\end{equation}
Here $X_{j,m}=X(k_j-k_m)$ are two-body scattering operators identical to
those for the $\delta$-function gas or the nearest-neighbor lattice model;
the eigenvectors and eigenvalues are given in S1.
The two-body phase shift $\theta_0 (k)$ is the phase-shift
for scattering from a $1/\sinh^2(r)$ potential, and is given as
\begin{equation}
\theta _0(k)=i\log \left[ {{{\Gamma (1+ik/ 2)\Gamma (\lambda -ik/
2)} \over {\Gamma (1-ik/ 2)\Gamma (\lambda +ik/ 2)}}} \right].
\end{equation}

Using the eigenvalues of the $X$'s from S1, one obtains coupled
equations for the $k$'s, the first being
\begin{equation}
Lk=2\pi I(k)+\sum\limits_{k} {\theta _0(k-k)}-\sum\limits_\alpha
{\theta (2(k-\alpha ))}.
\end{equation}
Here $\theta(k)$ is the phase-shift for a $\delta$-function gas:
$\theta(k) = 2\arctan(k/2\lambda)$.
The $I(k)$ are the quantum numbers, from $\log(\pm 1)/2\pi$.  The remaining
equations are identical with those of the nearest-neighbor model;
the $k$'s couple directly only to the second equation, which has the form
\begin{equation}
0=2\pi J(\alpha )-\sum\limits_k {\theta (2(\alpha -k))}+\ldots.
\label{eq-7}
\end{equation}
The remaining terms and equations are the same as for the nearest-neighbor
model, and their form and number depend upon what types of particles we
have.

%
%

\section{The Limit $\lambda\rightarrow\infty$}

When we take the limit $\lambda\rightarrow\infty$, we must also rescale
the asymptotic momenta $k$, defining a new variable $x\equiv k/2\lambda$.
Likewise the asymptotic momenta $\alpha$ for the compositional degrees of
freedom must be rescaled, but we use the same symbol, so
$\alpha\rightarrow\alpha/2\lambda$.  Then the phase shift for
particle-particle scattering has an expansion in $1/\lambda$ of the form
\begin{equation}
	\theta_0(k) \rightarrow 2\lambda \theta_0(x) + \theta_1(x) + \ldots,
\end{equation}
where
\begin{mathletters}
\begin{eqnarray}
\theta_0(x)
 &= &{1 \over 2}\left[ {x\log (1+1/ x^2)+i\log [(1-ix)/ (1+ix)]} \right], \\
\theta_0'(x)
 &= &{1 \over 2}\log (1+1/ x^2), \\
\theta_0''(x)
 &= &-{1 \over x}+{x \over {1+x^2}}.
\end{eqnarray}
\end{mathletters}
Finally, for one more redefinition, let  $\theta(x)\rightarrow 2\arctan(x)$.

We now expand the first equation in $1/\lambda$, obtaining to order $\lambda$
--- which we call the zeroth order, since it starts the expansion ---
the equation
\begin{equation}
\Lambda x=\sum\limits_{x} {\theta_0(x-x)}.
\label{eq-10}
\end{equation}
Let $N\rho(x) dx$ be the number of $x$'s in $x\rightarrow x+dx$.
Then the equation for $\rho$ becomes an integral equation, and upon
differentiating
\begin{equation}
{1 \over {2\pi d}}={1 \over {2\pi }}\int_{-A}^A {\theta_0'(x-x)}\rho
\label{eq-11}
(x)dx\equiv L\rho.
\end{equation}
Let $L^{-1}$ be the inverse of $L$, so $\rho=L^{-1} [1/2\pi d]$.
The normalization of $\rho$ is
\begin{equation}
1=\int_{-A}^A {\rho (x)dx}\equiv \openone^{\dagger} \rho,
\label{eq-12}
\end{equation}
while the classical ground state energy is
\begin{equation}
{E \over {4\lambda ^2N}}
=      {1 \over 2}\int_{-A}^A {x^2\rho (x)dx}
\equiv (x^2/ 2)^{\dagger} \rho
=      {1 \over 4}\sum\limits_{j=1}^\infty  {{1 \over {\sinh^2(j/d)}}}.
\label{eq-13}
\end{equation}
Before discussing the solution to this integral equation, we go on and
examine the first order corrections to the equation for the $x$'s.

%
%

\section{The First Order Equation}

We write the $x$'s which satisfy the zeroth order Eq.~(\ref{eq-10}) as $x_0$,
and then look for corrections to the $x$'s as
$x = x_0 + \delta x/2\lambda + \ldots$.
Let $\gamma(x) = \delta x(x) \rho(x)$, and
$N R(\alpha)d\alpha = \mbox{number of $\alpha$'s in $d\alpha$}$.
Then the first order equation is
\begin{equation}
L\gamma ={I(x) \over N}+{1 \over {2\pi }}\int_{-A}^A {\theta_1'(x-x)}\rho(x)dx-
{1 \over {2\pi }}\int_{-B}^B {\theta (2(x-\alpha )})R(\alpha )d\alpha.
\label{eq-14}
\end{equation}
Now this equation is linear, so we can write $\gamma = \gamma_0 + \gamma_1$,
where $\gamma_0$ is the correction for the elastic modes, while
$\gamma_1$ is the correction from the compositional modes.
We are interested only in the compositional degrees of freedom in this Letter,
so we write $\gamma$ for $\gamma_1$, which obeys the integral equation
\begin{equation}
L\gamma =-{1 \over {2\pi }}\int_{-B}^B {\theta (2(x-\alpha )})R(\alpha
)d\alpha \equiv -g.
\label{eq-15}
\end{equation}
Expanding the energy, and momentum to the same order, we find
\begin{mathletters}
\label{eq-16}
\begin{eqnarray}
\Delta E/ 2\lambda N
&= &x^{\dagger} \gamma
 =-x^{\dagger} L^{-1}g
 =-g^{\dagger} L^{-1}x
 =-g^{\dagger} e
 =e^{\dagger} g
 =E_{latt}/ N,\\
\Delta P/ dN
&= &1^{\dagger} \gamma / d
 =-1^{\dagger} L^{-1}g/ d
 =-g^{\dagger} L^{-1}1/ d
 =-2\pi g^{\dagger} \rho
 =P_{latt}/ N.
\end{eqnarray}
\end{mathletters}
Here, $e$ and the derivative $e'$ obey the equations
\begin{equation}
Le=x,\quad \mbox{and} \quad Le=x^2/ 2-\mu ,\;e(\pm A)=0.
\label{eq-17}
\end{equation}
Remember, all integrals over $x$ are integrals only over the interval $-A$ to
$A$.
These expressions of Eq.~(\ref{eq-16}) are exactly the energy and momentum for
the lattice Hamiltonian $H_{latt}$ of Eq.~(\ref{eq-2}).
The classical ground state energy in Eq.~(\ref{eq-13}) can also be expressed as
$E/4\lambda^2 N = \openone^{\dagger} e/2\pi d + \mu$.

As shown in Eq.~(\ref{eq-7}), the only effect of the $x$'s --- and thus the
$k$'s --- on the $\alpha$'s is through the expression
\begin{equation}
\sum\limits_x {\theta (2(\alpha -x))}=N\int_{-A}^A {\theta (2(\alpha
-x))}\rho (x)dx\equiv Np(\alpha ).
\label{eq-18}
\end{equation}
Returning to the expressions for the energy and momentum of the lattice
Hamiltonian Eq.~(\ref{eq-16}), and substituting the expression of
Eq.~(\ref{eq-15})
for $g$, we see that they can be written as
\begin{mathletters}
\label{eq-19}
\begin{eqnarray}
P_{latt}/ N
&= &\int_{-B}^B {p(\alpha )}R(\alpha )d\alpha =p^{\dagger} BR,\\
E_{latt}/ N
&= &\int_{-B}^B {\varepsilon (\alpha )}R(\alpha )d\alpha
 =\varepsilon ^{\dagger} BR,
\end{eqnarray}
\end{mathletters}
with $\varepsilon$ defined by
\begin{equation}
\varepsilon (\alpha )\equiv {1 \over \pi }\int_{-A}^A {\theta '(2(\alpha
-x))}e(x)dx.
\label{eq-20}
\end{equation}
Eliminating $\alpha$ between $p(\alpha)$ and $\varepsilon(\alpha)$ gives the
dispersion
relation $\varepsilon(p)$ for a single excitation of the lattice Hamiltonian.
This can also be expressed by the Fourier transform of the hopping matrix
element, and so
\begin{equation}
\varepsilon (p)=-\sum\limits_{j=1}^\infty  {{{1-(-1)^j\cos (jp)} \over
{\sinh ^2(j/ d)}}}.
\end{equation}
This dispersion curve is shown in Fig.~(\ref{fig-1}).

%
%

\section{The Densities \protect$\rho(\lowercase{x})$ and
\protect$\lowercase{e}(\lowercase{x})$}

As will become clearer, the solution to the hyperbolic lattice problem
depends crucially on the two classical ground state densities $\rho(x)$
and $e(x)$, obeying Eq.~(\ref{eq-11}), (\ref{eq-12}) and (\ref{eq-17}).
Unfortunately, these equations cannot be solved in closed form, even though
we know the densities obey an infinite number of sum rules, arising from the
classical limit of the Lax equations.
The energy sum rule of Eq.~(\ref{eq-13}) is the first of this heirarchy.

However, we can derive what appears to be a convergent expansion for
$\rho(x)$ and $e(x)$ in terms of the Chebyshev $T$ and $U$ polynomials,
of the form
\begin{mathletters}
\begin{eqnarray}
\rho (x)
&= &{1 \over {\pi \sqrt {A^2-x^2}}}
 \left[
 {1+\sum\limits_{j=1}^\infty  {\rho _jT_{2j}(x/ A)}}
 \right],\\
e(x)
&= &-2\sqrt {A^2-x^2}\sum\limits_{j=0}^\infty  {e_jU_{2j}(x/ A)}.
\end{eqnarray}
\end{mathletters}

In the nearest-neighbor limit of $A\rightarrow 0$, when
$\theta_0'(x)\to -\log |x|$,
one finds that the expansions stop at the first term, so that
$\rho (x)\to 1/ \pi \sqrt {A^2-x^2}\approx \delta (x)$,
and
$e(x)\to -2\sqrt {A^2-x^2}\approx -\pi A^2\delta (x)$.
This gives $d^{-1}\rightarrow -\log [A/2]$.

At the other limit of an inverse square interaction, when
$A\rightarrow\infty$, then
$\theta_0'(x)\to \pi \delta (x)$,
so $\rho(x)\rightarrow 1/2A$, $d\rightarrow 2A/\pi$, and
$e\rightarrow x^2-2\mu=x^2-A^2$.  In Fig.~(\ref{fig-2}) and (\ref{fig-3})
we show $\rho(x)$ and $e(x)$ for representative values of $A$.

%
%

\section{Summary of Results for the Nearest-Neighbor Model}
\label{sec-6}

We make extensive use of the solution from S1 for the nearest-neighbor
chain, so we summarize the results here.  We restrict ourself in this
letter to the most interesting case of the $F^f$ system, with $N_j$ particles
of type $j$, and $N_1\geq\ldots\geq N_f$.
Let $M_1=N-N_1$, $M_2=N-N_1-N_2$, $\ldots$, $M_f-1=N_f$.
Then there are $f-1$ coupled equations for $f-1$ sets of roots, the $j$th set
being $M_j$ roots in number.
For the ground state in the thermodynamic limit, these sets of roots distribute
with densities $r_j(\alpha)$ between the limits $-b_j$ and $b_j$, normalized
so that
\begin{equation}
\int_{-b_j}^{b_j} {r_j(\alpha )d\alpha} =M_j/ N\equiv
m_j,\quad \mbox{or}\quad\openone^{\dagger} br=m.
\end{equation}
We have adopted a very useful notation where the $f-1$ $r$'s and $m$'s are
arranged as column vectors.  Let $K_n$ represent an integral operator with
difference kernel
\begin{equation}
K_n(\alpha )={1 \over {2\pi }}{{2n} \over {1+n^2\alpha ^2}}=\theta'(n\alpha )/
2\pi.
\end{equation}
Finally, let $\xi$ be a column vector with components $x_j= \delta_{j1}
K_2(\alpha)$,
$b$ a matrix projection operator which imposes the integration limits
$\pm b_j$ on $\alpha_j$, and $K$ a symmetric matrix of integral operators with
elements $K_{j,m}=\delta_{j,m} K_2 - (\delta_{j,m-1}+\delta_{j,m+1} )K_1$.
Then the ground state densities satisfy the $f-1$ coupled integral
equations  $\xi = r + K b r = (\openone+K) b r$.
With the Hamiltonian normalized as
\begin{equation}
H_{nn}=-\sum\limits_{j>k} {[1+P_{jk}]},
\end{equation}
the energy is $E/N=-2\pi \xi^{\dagger} b r$.
Finally, the momentum is $P/N= \zeta^{\dagger} br$, with
$\zeta' \equiv 2\pi \xi$, which of course is zero for the ground state.

In the case of the absolute ground state, when there are equal numbers
of each type of particle so $n_j=N_j/N=1/f$, then all limits $b_j=\infty$,
so $b=\openone$, and the equation can be solved by Fourier transforms.
The eigenvalues of the integral operators $K_n$ are given by the Fourier
transform of the kernel,
\begin{equation}
\tilde K_n(s)\equiv \int_{-\infty }^\infty  {e^{-is\alpha }K_n(\alpha
)d\alpha }=e^{-|s|/ n}.
\end{equation}
We see $\tilde K_1=(\tilde K_2)^2$.
We now define the resolvent operator $J$ by
$(\openone+J)(\openone+K)=(\openone+K)(\openone+J)=\openone$.
Then $\openone+J$ is also a symmetric matrix of integral operators with
difference kernels, whose eigenvalues are given by the Fourier transforms
\begin{equation}
[\tilde{\openone}+\tilde J]_{j,m}=e^{|s|/ 2}{{\sinh (s(f-j)/ 2)\sinh (sm/ 2)}
\over {\sinh (fs/ 2)\sinh (s/ 2)}},\;j\ge m.
\end{equation}
The Fourier transform of the densities are then given as
\begin{equation}
\tilde r_j(s)={{\sinh (s(f-j)/ 2)} \over {\sinh (fs/ 2)}},
\end{equation}
confirming $m_j=\tilde r_j(0)=1-j/ f$
so $n_j = 1/f$.  The ground state energy is calculated by Parsival's theorem
as
\begin{equation}
E/ N=-\int_{-\infty }^\infty  {e^{-|s|/ 2}\tilde r_1(s)ds}=-2[\psi (1)-\psi
(1/ f)]/ f,
\end{equation}
with $\psi(x)$ as the digamma function.

For the excitations about the absolute ground state, there are $f-1$
branches, each of which is excited pair-wise, as a particle and a hole.  The
energy and momentum for these branches are given as
\begin{mathletters}
\label{eq-30}
\begin{eqnarray}
\Delta E_j(\alpha )
 &= &{{2\pi } \over f}{{\sin (\pi j/ f)} \over
     {\cosh (2\pi \alpha / f)-\cos (\pi j/ f)}}, \\
\Delta P_j(\alpha )
 &= &2\arctan [\cot (\pi j/ 2f)\tanh (\pi \alpha / f)]-\pi (1-j/ f).
\end{eqnarray}
\end{mathletters}
Eliminating $\alpha$ between the two equations gives a dispersion relation
\begin{equation}
\omega _j(k)=2\pi [\cos (\pi j/ f-|k|)-\cos (\pi j/ f)]/ \sin (\pi j/f),
\end{equation}
for $|k|\leq 2\pi j/f$; it is periodic with period $2\pi j/f$.  At $k=0$, all
branches have a common velocity $v=2\pi /f$.  This leads to our final result,
that if we deviate slightly from equal filling, so the concentrations are
$n_j = 1/f + \delta n_j$, then to leading order in the $\delta n$'s,
\begin{equation}
\Delta E/N \simeq \frac{\pi^2}{f} \sum_{j=1}^{f} \delta n_j^2,
\end{equation}
and thus the susceptibility exists and is isotropic.

%
%

\section{Results for the Hyperbolic Model}

We shall make extensive use of the solution to the nearest-neighbor
problem to construct a solution to the general hyperbolic problem, so let us
rewrite the coupled equations for the general problem as
$\hat \xi =(\openone +K)b\hat r$.
The inhomogeneous term is given by Eq.~(\ref{eq-18}), and can be rewritten as
$\hat \xi =\xi \rho $, by defining $\xi$ as an a column vector of integral
operators, with $\xi_j = \delta_{1j} K_2$.
This is a natural extension of the previous notation.  Since the equations are
linear, the solution to the general problem is given by superposition as
$\hat r=r\rho $, $r$ being the solution to the nearest-neighbor problem.
For the lattice with equal filling, when $b=\openone$, $r$ is just a vector of
translationally invariant operators.

Examining the Eqs.~(\ref{eq-18}), (\ref{eq-19}) and (\ref{eq-20}) giving the
energy
and momentum, we see that they can be rewritten as
\begin{mathletters}
\label{eq-33}
\begin{eqnarray}
P_{latt}/ N
 &= &\rho ^{\dagger} \zeta ^{\dagger} b\hat r=\rho ^{\dagger} \zeta ^{\dagger}
br\rho , \\
E_{latt}/ N
 &= &e^{\dagger} \xi ^{\dagger} b\hat r=e^{\dagger} \xi ^{\dagger} br\rho ,
\end{eqnarray}
\end{mathletters}
and so both can be expressed as quadratic forms in $\rho$ and $e$.
Here $\zeta'\equiv 2\pi \xi$.
The concentrations $n_j$ remain unchanged, due to the normalization of $\rho$.
We see that singularities in the groundstate only occur at surfaces of equal
filling, the same as for the nearest-neighbor case.  Again, taking the absolute
ground state with equal filling, we can explicitly evaluate these expressions
using the results of section \ref{sec-6}, giving for the ground state energy
\begin{equation}
E_{latt}/ N= -\int_{-A}^A {e(x)dx}\int_{-A}^A {\rho (x')dx'}
\mbox{Re} [\psi (1/ f+i(x-x')/ f)-\psi (1+i(x-x')/ f)]/ \pi f.
\end{equation}
The ground state momentum is of course zero.  In Fig.~(\ref{fig-4}) we show the
ground state energy as a function of the number of species of fermions $f$, for
several typical densities $d$, and hence various ranges for the hyperbolic
interaction.

For the low-lying excitations, one calculates the shift in the asymptotic
momenta, weighted by the ground state distributions, and this quantity does
not depend upon the momentum function $p(\alpha)$.
Thus, the only change for the general case comes from the expressions
Eq.~(\ref{eq-33}) for the energy and momentum, so we can write for the
excitations
\begin{mathletters}
\begin{eqnarray}
\Delta P_{latt}
 &= &\int_{-A}^A {\Delta P_j(\alpha -x)\rho(x)dx}, \\
\Delta E_{latt}
 &= &-\int_{-A}^A {\Delta E_j(\alpha -x)e(x)dx}/ 2\pi ,
\end{eqnarray}
\end{mathletters}
where $\Delta P_j(\alpha)$ and $\Delta E_j(\alpha)$ are the nearest-neighbor
expressions of Eq.~(\ref{eq-30}). The corresponding hydrodynamic velocities for
these gapless excitations are all identical, and equal to
\begin{equation}
v={{\int_{-A}^A {\exp (2\pi x/ f)|e(x)|dx}} \over {f\int_{-A}^A {\exp
(2\pi x/ f)\rho (x)dx}}}.
\label{eq-36}
\end{equation}
This expression is very much like that for a recently solved model for a
magnetic fluid, which has the Heisenberg-Ising model imbedded in a
continuum system \cite{9}.  In Fig.~(\ref{fig-5}) we show the hydrodynamic
velocity $v$ as a function of the number of species of fermions $f$, for
several
typical densities $d$, and hence various ranges for the hyperbolic interaction.

The susceptibility can be written using the expression Eq.~(\ref{eq-36}) for
$v$
as
\begin{equation}
\Delta E/ N\approx {{\pi v} \over 2}\sum\limits_{j=1}^f {\delta
n_j^2}\equiv {{\chi ^{-1}} \over 2}\sum\limits_{j=1}^f {\delta n_j^2}.
\end{equation}
Finally, by a thermodynamic argument \cite{10}, if we introduce fluxes
$\Phi_j = N \phi_j$ conjugate to the charge of type $j$, then
\begin{equation}
\Delta E/ N\approx {v \over {2\pi }}\sum\limits_{j=1}^f {\delta \phi
_j^2}\equiv {D \over 2}\sum\limits_{j=1}^f {\delta \phi _j^2},
\end{equation}
so the charge stiffness $D$ obeys $D \chi^{-1} = v^2$.

\acknowledgements
We would like to thank Joel Campbell for his help in solving for $\rho(x)$
and $e(x)$.
R.A.R.\ gratefully acknowledges partial support by the Germanistic Society
of America.


\begin{figure}
  \caption{The dispersion curve $e(p)$ is shown for various values of the
 density.
	The energy is measured in terms of the nearest-neighbor interaction
	energy $\sinh^{-2}(1/d)$.  In all our figures, we have taken as examples,
	limits of $A=1/2, 1, 2, 4$ corresponding to densities of $d = 0.6735,
	1.0388, 1.7293, 3.0596$ respectively.  In addition, we have the limit
	$A\rightarrow 0$, $d\rightarrow 0$, corresponding to the nearest-neighbor
 interaction, and the limit $A\rightarrow\infty$, $d\rightarrow\infty$,
 corresponding to the $1/r^2$ interaction.
  \label{fig-1}}
\end{figure}

\begin{figure}
  \caption{The classical ground state distribution function $\rho(x)$ is shown
 for representative values of the parameter $A$.
  \label{fig-2}}
\end{figure}

\begin{figure}
  \caption{The classical ground state distribution function $e(x)$ is shown for
	representative values of the parameter $A$.  It is measured in terms of
	the nearest-neighbor interaction energy $\sinh^{-2}(1/d)$.
  \label{fig-3}}
\end{figure}

\begin{figure}
  \caption{The ground state energy $E_{latt}/N$ is shown as a function of the
 number of species of fermions $f$, for representative values of the parameter
 $A$.
	It is measured in terms of the nearest-neighbor interaction energy
	$\sinh^{-2}(1/d)$.
  \label{fig-4}}
\end{figure}

\begin{figure}
  \caption{The hydrodynamic velocity $v$ is shown as a function of the number
of
	species of fermions $f$, for representative values of the parameter $A$.  It
is
	measured in terms of the nearest-neighbor interaction energy
	$\sinh^{-2}(1/d)$.
  \label{fig-5}}
\end{figure}

\end{document}